%% file: ckm2010.tex
\documentclass[12pt]{article}
\usepackage{graphicx}
\usepackage[latin1]{inputenc}
\usepackage{units}
\usepackage{amssymb}

\newcommand{\Vud}{$V_\mathrm{ud}$\ }

\newcommand\pubnumber{} 
\newcommand\pubdate{May 27, 2011}

\def\heidelberg{Physikalisches Institut, Universit{\"a}t Heidelberg, Philosophenweg~12, 69120~Heidelberg, Germany}
\def\ill{Institut Laue-Langevin\footnote{Present address: \heidelberg}\\ 6, rue Jules Horowitz,
38042~Grenoble Cedex~9, France}
\def\support{\footnote{Work supported by the priority programme SPP~1491 of the German Research Foundation~(DFG).}}

\def\Title#1{\begin{center} {\Large #1 } \end{center}}
\def\Author#1{\begin{center}{ \sc #1} \end{center}}
\def\Address#1{\begin{center}{ \it #1} \end{center}}

\newcommand\pubblock{\rightline{\begin{tabular}{l} \pubnumber\\
         \pubdate  \end{tabular}}}
\newenvironment{Abstract}{\begin{quotation}  }{\end{quotation}}
\newenvironment{Presented}{\begin{quotation} \begin{center} 
             PRESENTED AT\end{center}\bigskip 
      \begin{center}\begin{large}}{\end{large}\end{center} \end{quotation}}
\def\Acknowledgements{\bigskip  \bigskip \begin{center} \begin{large}
             \bf ACKNOWLEDGEMENTS \end{large}\end{center}}

\input econfmacros.tex

\begin{document}
\begin{titlepage}
\pubblock

\vfill
\Title{Experimental Status of \Vud from Neutron Decay}
\vfill
\Author{Bastian~M{\"a}rkisch\support}
\Address{\ill}
\vfill
\begin{Abstract}
The matrix element \Vud of the CKM matrix can be determined by two independent measurements on neutron decay: the neutron lifetime $\tau_n$ and the ratio of coupling constants $\lambda=g_A/g_V$, which is most precisely determined by measurements of the beta asymmetry angular correlation coefficient~$A$. We present recent progress on the determination of these values and derive a world average of $V_\mathrm{ud} = 0.9743 (2)_\mathrm{RC} (8)_{\tau_n} (12)_{\lambda}$.
\end{Abstract}
\vfill
\begin{Presented}
CKM2010, the 6th International Workshop on the CKM Unitarity Triangle, \\ University of Warwick, UK, \\ 6-10 September 2010\\
\end{Presented}
\vfill
\end{titlepage}
\def\thefootnote{\fnsymbol{footnote}}
\setcounter{footnote}{0}

\section{Introduction}

To obtain the matrix element \Vud from the decay of the free neutron only two separate inputs are required. These are the neutron lifetime $\tau_n$ and the ratio of axial vector and vector coupling constants $\lambda$, which can be determined by measurements of angular correlations in neutron decay \cite{Jackson57,Wilkinson82}. \Vud can then be determined by \cite{Marciano06}
\begin{equation} 
\label{eq:Vud}
\left|V_\mathrm{ud}\right|^2 = \frac{(4908.7 \pm 1.9)s}{\tau_n \left(1+3\lambda^2 \right)}, 
\end{equation}
where the numerator includes all constants, with the Fermi coupling constant precisely determined in muon decay, and the theoretical uncertainty of the radiative corrections (RC) (see also \cite{Marciano10}).

\section{Neutron Lifetime}

Neutron lifetime experiments fall into two separate groups. The first group can be classified as \emph{``counting the dead''}: a beam of neutrons passes an active volume of length~$l$. Electrons or protons from neutron decay from within that volume are counted. An absolute measurement is made of the thermal equivalent flux $\phi_\mathrm{c}$ of the neutron beam. Typically, the decay probability of the neutron within the volume is on the order of $10^{-6}$. The neutron lifetime $\tau_n$ is then derived from the electron or proton count rate~$r$:
\begin{equation}
r \approx  \frac{\phi_\mathrm{c} \, l}{v_0 \, \tau_n}, \quad \mathrm{where} \quad v_0 = 2200\,\unit{m/ s}
\end{equation}
Experimentally challenging are several absolute measurements that are required: the neutron flux, the electron/proton count rate and the length of the active volume. 

Figure \ref{fig:lifetime1} gives an overview of all neutron lifetime experiments with a precision better than $10\,\unit{s}$. The \emph{in-beam} measurements are indicated in blue. The most precise experiment of this kind is carried out at NIST \cite{Nico05}. From this type of measurement we obtain an average of
\begin{equation}
\tau_n = 887.6 \pm 2.7\,\unit{s} \qquad \mbox{(in-beam measurements only)}.
\end{equation}

The second group of experiments uses the property of ultracold neutrons (UCN) to be reflected from surfaces under any angle of incidence. UCNs are trapped in a closed vessel and the their lifetime is measured by \emph{``counting the survivors''} $N(t)$ after some storage time $t$:
\begin{equation}
\frac{1}{\tau_n} = \frac{1}{t} \ln \frac{N(0)}{N(t)}
\end{equation}
The reflection probability on the trap surfaces is typically on the order of ($1 - 10^{-5}$) per collision, which causes additional, neutron energy dependent losses. These are accounted for by varying the trap size and extrapolation to an infinite volume. Figure \ref{fig:lifetime2} shows the difference between the longest neutron storage time achieved and the derived neutron lifetime. We note that the measurement with the smallest extrapolation \cite{Serebrov05} is currently not included by the PDG in the world average of the neutron lifetime \cite{PDG2010}.

Since the temperature of the trap wall is typically much larger ($\approx 100\,\unit{K}$) than the temperature of the UCNs ($\approx 1\,\unit{mK}$), energy transfer to the UCNs is possible. The most recently published Mambo~II experiment \cite{Pichlmaier10} is the improved successor of the original Mambo experiment \cite{Mampe89}. Their result is -- to leading order -- insensitive to UCN spectrum changes due to careful preparation of the initial spectrum and clever choice of measurement time intervals.

\begin{figure}[ht]
\begin{minipage}[t]{0.48\linewidth}
\centering
\scriptsize
\input{lifetime2.tex}
\caption{History of neutron lifetime measurements with a precision better than $10\,\unit{s}$: in-beam measurements are indicated in blue, storage experiments in green. The results of \cite{Serebrov05} and \cite{Pichlmaier10} are not included in the PDG~2010 average.}
\label{fig:lifetime1}
\end{minipage}
\hfill
\begin{minipage}[t]{0.48\linewidth}
\centering
\scriptsize
\input{lifetime3.tex}
\caption{Neutron lifetime determination by storage of ultracold neutrons: The largest storage time achieved is indicated along with the derived neutron lifetime.}
\label{fig:lifetime2}
\end{minipage}
\end{figure}

Like the authors of \cite{Pichlmaier10,Konrad10,Dubbers11} we propose to include \emph{all} measurements \cite{Spivak88,Mampe89,Nesvi92,Mampe93,Byrne96,Arzumanov00,Nico05,Serebrov05,Pichlmaier10} in the average and scale the error according to PDG procedures, despite the large discrepancy between the measurements with the smallest quoted uncertainty \cite{Arzumanov00} and \cite{Serebrov05}. We obtain:
\begin{equation}
\label{eq:lifetime}
\tau_n = 881.8 \pm 1.4 \,\unit{s}, \qquad S = 2.7,  \qquad \mbox{(all measurements),}
\end{equation}
where $S$ denotes the scaling factor. We note that the authors of \cite{Serebrov10} have published Monte-Carlo studies of the setups used for \cite{Mampe89} and \cite{Arzumanov00} where they report large systematic errors.

Further experimental details on neutron lifetime experiments can be found in the reviews \cite{Paul09,Abele08,Nico09,Dubbers11}.

\section{Ratio of Coupling Constants $\lambda$}

Assuming vector current conservation, the axial vector coupling constant can be determined within the Standard Model from a variety of angular correlation measurements \cite{Jackson57, Wilkinson82}. The most precise determination comes from measurements of the beta asymmetry $A$ correlation coefficient, which describes the correlation between neutron spin and electron momentum. To leading order $A_0$ this asymmetry is given by
\begin{equation}
A_0 = \frac{-2 \left( \lambda^2 - |\lambda| \right)}{1 + 3 \lambda^2},
\end{equation}
where $\lambda = g_A/g_V$ is the ratio of axial vector and vector coupling constants. Due to the equally high sensitivity on $\lambda$, the determination of the electron-neutrino angular correlation $a$ is another candidate. 

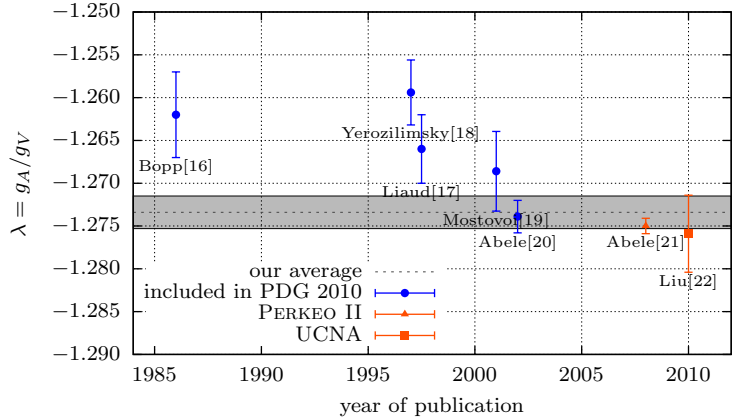
\begin{figure}
\centering
\scriptsize
\input{lambda.tex}
\caption{\label{fig:lambda}Determination of $\lambda$ from correlation measurements. Included are the preliminary value by the \textsc{Perkeo~II} collaboration \cite{Abele08}, which combines \cite{Abele02} and \cite{Mund11}, and the recent result of the UCNA experiment \cite{Liu10}. The band indicates the proposed updated average.}
\end{figure}

The most recent determinations of the beta asymmetry have been performed using a similar scheme as the original 
\textsc{Perkeo} instrument \cite{Bopp86}. The decay of polarized neutrons in a strong magnetic field is analysed by electron spectroscopy with a solid angle coverage of $2 \times 2\pi$. The \textsc{Perkeo~II} collaboration \cite{Abele02,Abele08} uses a split coil setup and a strong cold neutron beam, whereas the UCNA collaboration uses ultracold neutrons stored within a strong solenoid \cite{Liu10}. In these experiments backscattering of electrons from the detectors, a serious source of error in $\beta$-spectroscopy, is strongly suppressed by a decrease in magnetic field strength towards the detectors and detection of backscattered electrons in the second $2\pi$ detector.
Only minor corrections at the percent level are required to the raw measured value, whereas previously corrections due to e.g. solid angle, neutron polarisation and magnetic mirror effects attributed to $15-29\%$. A history plot including corrections is found in \cite{Abele08}.

Figure \ref{fig:lambda} gives an overview of recent precise measurements of $\lambda$. We combine the results of \cite{Bopp86,Liaud97,Yero97} (all included in the PDG~2010 analysis) and the new results of \cite{Abele08,Liu10} to obtain
\begin{equation}
\label{eq:lambda}
\lambda = -1.2734(19),\qquad S=2.3,
\end{equation}
were we have scaled the error bar according to standard PDG procedures.

Several new projects are currently running or being built to improve the experimental accuracy. The \textsc{Perkeo III} spectrometer \cite{Maerkisch09} increases the size of the active volume by almost two orders of magnitude, which makes it feasible to use a pulsed neutron beam. Beam related background can now be measured under the same conditions as the decay signal itself and can thus be fully subtracted. Data of a run in 2009 is currently being analysed.

The instrument PERC \cite{Dubbers08}, which is under development by an international collaboration, follows a radically novel concept to improve systematics and statistics: to maximise the phase space of the neutrons, a neutron guide consisting of non-depolarising mirrors in a strong magnetic field is used as \emph{active volume}. A magnetic filter is then used to limit the phase space of the emerging electrons and protons. A detailed analysis shows that all relevant sources of systematic error can be controlled on the $10^{-4}$ level or better, an improvement by one order of magnitude in comparison to existing spectrometers.

Complementary to electron asymmetry measurements, several projects (\emph{a}SPECT, aCORN, Nab, PERC) aim to derive the ratio of coupling constants $\lambda$ with competitive precision from the electron-neutrino correlation coefficient $a$.

\section{Summary and Outlook}

Combining our new averages (\ref{eq:lifetime}) and (\ref{eq:lambda}) with equation (\ref{eq:Vud}), we obtain from neutron decay data alone
\begin{equation}
V_\mathrm{ud} = 0.9743 \,(2)_\mathrm{RC} \,(8)_{\tau_n} \,(12)_\lambda.
\label{eq:Vudvalue}
\end{equation}
Due to the discrepancies in the published values, the error of the neutron lifetime $\tau_n$ has been scaled by $2.7$ and the error on the ratio of coupling constants $\lambda$ has been scaled by $2.3$. 

This value is in perfect agreement with the value $V_\mathrm{ud} = 0.97425 \,(21)_\mathrm{RC} \,(8)_\mathrm{exp}$ from nuclear beta decays (see \cite{Hardy09,Towner10,Malconian10}), albeit with a much larger experimental error. Unlike nuclear beta decay, current neutron beta decay experiments are not limited by the theoretical knowledge of nuclear corrections.

We note that the PDG~2010 finds $V_\mathrm{ud} = 0.9746 \,(2)_\mathrm{RC} \,(4)_{\tau_n} \,(18)_\lambda$ \cite{PDG2010} based on a subset of the data used in our analysis.

New experiments on the neutron lifetime using novel techniques such as wall-less magnetic storage of neutrons and \emph{in situ} observation of decay protons or electrons will significantly reduce systematic uncertainties and are expected to resolve the current disagreements. Several new measurements have been proposed e.g. at a recent conference at the ILL \cite{Soldner09} and some of them have already collected data.

The angular correlation measurements are also pursued by a lively community. Currently, the  \textsc{Perkeo~III}, UCNA, \textit{a}{SPECT} and aCorn experiments are running or analysing data. The next generation projects PERC and Nab aim to improve the current precision to the same level as in nuclear beta decay.

\Acknowledgements

The author would like to thank the organizers and conveners for their kind invitation to this inspiring conference. The author acknowledges fruitful discussions on the details of experiments with many of his colleagues at the Institut Laue-Langevin and at the Universities of Heidelberg and Vienna.

\end{document}

%% file: econfmacros.tex
\def\beq{\begin{equation}}
\def\eeq#1{\label{#1}\end{equation}}
\def\eeqn{\end{equation}}

\def\beqa{\begin{eqnarray}}
\def\eeqa#1{\label{#1}\end{eqnarray}}
\def\eeqan{\end{eqnarray}}

\let\bar=\overbar

\def\Dslash{\not{\hbox{\kern-4pt $D$}}}
\def\dslash{\not{\hbox{\kern-2pt $\del$}}}

\def\msb{{\bar{\ssstyle M \kern -1pt S}}}



%% file: lifetime2.tex
\begingroup
  \makeatletter
  \providecommand\color[2][]{%
    \GenericError{(gnuplot) \space\space\space\@spaces}{%
      Package color not loaded in conjunction with
      terminal option `colourtext'%
    }{See the gnuplot documentation for explanation.%
    }{Either use 'blacktext' in gnuplot or load the package
      color.sty in LaTeX.}%
    \renewcommand\color[2][]{}%
  }%
  \providecommand\includegraphics[2][]{%
    \GenericError{(gnuplot) \space\space\space\@spaces}{%
      Package graphicx or graphics not loaded%
    }{See the gnuplot documentation for explanation.%
    }{The gnuplot epslatex terminal needs graphicx.sty or graphics.sty.}%
    \renewcommand\includegraphics[2][]{}%
  }%
  \providecommand\rotatebox[2]{#2}%
  \@ifundefined{ifGPcolor}{%
    \newif\ifGPcolor
    \GPcolortrue
  }{}%
  \@ifundefined{ifGPblacktext}{%
    \newif\ifGPblacktext
    \GPblacktexttrue
  }{}%
  \let\gplgaddtomacro\g@addto@macro
  \gdef\gplbacktext{}%
  \gdef\gplfronttext{}%
  \makeatother
  \ifGPblacktext
    \def\colorrgb#1{}%
    \def\colorgray#1{}%
  \else
    \ifGPcolor
      \def\colorrgb#1{\color[rgb]{#1}}%
      \def\colorgray#1{\color[gray]{#1}}%
      \expandafter\def\csname LTw\endcsname{\color{white}}%
      \expandafter\def\csname LTb\endcsname{\color{black}}%
      \expandafter\def\csname LTa\endcsname{\color{black}}%
      \expandafter\def\csname LT0\endcsname{\color[rgb]{1,0,0}}%
      \expandafter\def\csname LT1\endcsname{\color[rgb]{0,1,0}}%
      \expandafter\def\csname LT2\endcsname{\color[rgb]{0,0,1}}%
      \expandafter\def\csname LT3\endcsname{\color[rgb]{1,0,1}}%
      \expandafter\def\csname LT4\endcsname{\color[rgb]{0,1,1}}%
      \expandafter\def\csname LT5\endcsname{\color[rgb]{1,1,0}}%
      \expandafter\def\csname LT6\endcsname{\color[rgb]{0,0,0}}%
      \expandafter\def\csname LT7\endcsname{\color[rgb]{1,0.3,0}}%
      \expandafter\def\csname LT8\endcsname{\color[rgb]{0.5,0.5,0.5}}%
    \else
      \def\colorrgb#1{\color{black}}%
      \def\colorgray#1{\color[gray]{#1}}%
      \expandafter\def\csname LTw\endcsname{\color{white}}%
      \expandafter\def\csname LTb\endcsname{\color{black}}%
      \expandafter\def\csname LTa\endcsname{\color{black}}%
      \expandafter\def\csname LT0\endcsname{\color{black}}%
      \expandafter\def\csname LT1\endcsname{\color{black}}%
      \expandafter\def\csname LT2\endcsname{\color{black}}%
      \expandafter\def\csname LT3\endcsname{\color{black}}%
      \expandafter\def\csname LT4\endcsname{\color{black}}%
      \expandafter\def\csname LT5\endcsname{\color{black}}%
      \expandafter\def\csname LT6\endcsname{\color{black}}%
      \expandafter\def\csname LT7\endcsname{\color{black}}%
      \expandafter\def\csname LT8\endcsname{\color{black}}%
    \fi
  \fi
  \setlength{\unitlength}{0.0500bp}%
  \begin{picture}(3854.00,3288.00)%
    \gplgaddtomacro\gplbacktext{%
      \csname LTb\endcsname%
      \put(384,512){\makebox(0,0)[r]{\strut{}$870$}}%
      \csname LTb\endcsname%
      \put(384,881){\makebox(0,0)[r]{\strut{}$875$}}%
      \csname LTb\endcsname%
      \put(384,1250){\makebox(0,0)[r]{\strut{}$880$}}%
      \csname LTb\endcsname%
      \put(384,1619){\makebox(0,0)[r]{\strut{}$885$}}%
      \csname LTb\endcsname%
      \put(384,1988){\makebox(0,0)[r]{\strut{}$890$}}%
      \csname LTb\endcsname%
      \put(384,2357){\makebox(0,0)[r]{\strut{}$895$}}%
      \csname LTb\endcsname%
      \put(384,2726){\makebox(0,0)[r]{\strut{}$900$}}%
      \csname LTb\endcsname%
      \put(384,3095){\makebox(0,0)[r]{\strut{}$905$}}%
      \csname LTb\endcsname%
      \put(962,352){\makebox(0,0){\strut{}$1990$}}%
      \csname LTb\endcsname%
      \put(1564,352){\makebox(0,0){\strut{}$1995$}}%
      \csname LTb\endcsname%
      \put(2166,352){\makebox(0,0){\strut{}$2000$}}%
      \csname LTb\endcsname%
      \put(2769,352){\makebox(0,0){\strut{}$2005$}}%
      \csname LTb\endcsname%
      \put(3371,352){\makebox(0,0){\strut{}$2010$}}%
      \put(16,1803){\rotatebox{-270}{\makebox(0,0){\strut{}lifetime (s)}}}%
      \put(2166,112){\makebox(0,0){\strut{}year of publication}}%
    }%
    \gplgaddtomacro\gplfronttext{%
      \csname LTb\endcsname%
      \put(3118,2952){\makebox(0,0)[r]{\strut{}our average}}%
      \csname LTb\endcsname%
      \put(3118,2792){\makebox(0,0)[r]{\strut{}in beam}}%
      \csname LTb\endcsname%
      \put(3118,2632){\makebox(0,0)[r]{\strut{}material trap}}%
      \csname LTb\endcsname%
      \put(2769,1315){\makebox(0,0){\strut{}\rotatebox{45}{\tiny Nico\cite{Nico05}}}}%
      \put(1685,1424){\makebox(0,0){\strut{}\rotatebox{45}{\tiny Byrne\cite{Byrne96}}}}%
      \put(721,1250){\makebox(0,0){\strut{}\rotatebox{45}{\tiny Spivak\cite{Spivak88}}}}%
      \csname LTb\endcsname%
      \put(3371,1097){\makebox(0,0){\strut{}\rotatebox{45}{\tiny Pichlmaier\cite{Pichlmaier10}}}}%
      \put(2769,1009){\makebox(0,0){\strut{}\rotatebox{45}{\tiny Serebrov\cite{Serebrov05}}}}%
      \put(2166,1502){\makebox(0,0){\strut{}\rotatebox{45}{\tiny Arzumanov\cite{Arzumanov00}}}}%
      \put(1323,1169){\makebox(0,0){\strut{}\rotatebox{45}{\tiny Mampe\cite{Mampe93}}}}%
      \put(1203,1553){\makebox(0,0){\strut{}\rotatebox{45}{\tiny Nesvizhevsky\cite{Nesvi92}}}}%
      \put(841,1516){\makebox(0,0){\strut{}\rotatebox{45}{\tiny Mampe\cite{Mampe89}}}}%
    }%
    \gplbacktext
    \put(0,0){\includegraphics{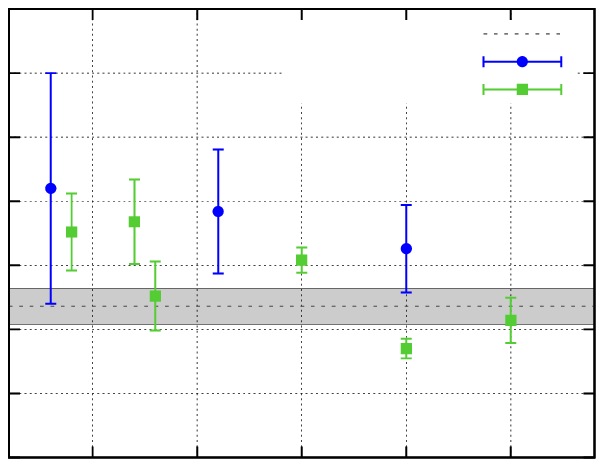}}%
    \gplfronttext
  \end{picture}%
\endgroup

%% file: lifetime3.tex
\begingroup
  \makeatletter
  \providecommand\color[2][]{%
    \GenericError{(gnuplot) \space\space\space\@spaces}{%
      Package color not loaded in conjunction with
      terminal option `colourtext'%
    }{See the gnuplot documentation for explanation.%
    }{Either use 'blacktext' in gnuplot or load the package
      color.sty in LaTeX.}%
    \renewcommand\color[2][]{}%
  }%
  \providecommand\includegraphics[2][]{%
    \GenericError{(gnuplot) \space\space\space\@spaces}{%
      Package graphicx or graphics not loaded%
    }{See the gnuplot documentation for explanation.%
    }{The gnuplot epslatex terminal needs graphicx.sty or graphics.sty.}%
    \renewcommand\includegraphics[2][]{}%
  }%
  \providecommand\rotatebox[2]{#2}%
  \@ifundefined{ifGPcolor}{%
    \newif\ifGPcolor
    \GPcolortrue
  }{}%
  \@ifundefined{ifGPblacktext}{%
    \newif\ifGPblacktext
    \GPblacktexttrue
  }{}%
  \let\gplgaddtomacro\g@addto@macro
  \gdef\gplbacktext{}%
  \gdef\gplfronttext{}%
  \makeatother
  \ifGPblacktext
    \def\colorrgb#1{}%
    \def\colorgray#1{}%
  \else
    \ifGPcolor
      \def\colorrgb#1{\color[rgb]{#1}}%
      \def\colorgray#1{\color[gray]{#1}}%
      \expandafter\def\csname LTw\endcsname{\color{white}}%
      \expandafter\def\csname LTb\endcsname{\color{black}}%
      \expandafter\def\csname LTa\endcsname{\color{black}}%
      \expandafter\def\csname LT0\endcsname{\color[rgb]{1,0,0}}%
      \expandafter\def\csname LT1\endcsname{\color[rgb]{0,1,0}}%
      \expandafter\def\csname LT2\endcsname{\color[rgb]{0,0,1}}%
      \expandafter\def\csname LT3\endcsname{\color[rgb]{1,0,1}}%
      \expandafter\def\csname LT4\endcsname{\color[rgb]{0,1,1}}%
      \expandafter\def\csname LT5\endcsname{\color[rgb]{1,1,0}}%
      \expandafter\def\csname LT6\endcsname{\color[rgb]{0,0,0}}%
      \expandafter\def\csname LT7\endcsname{\color[rgb]{1,0.3,0}}%
      \expandafter\def\csname LT8\endcsname{\color[rgb]{0.5,0.5,0.5}}%
    \else
      \def\colorrgb#1{\color{black}}%
      \def\colorgray#1{\color[gray]{#1}}%
      \expandafter\def\csname LTw\endcsname{\color{white}}%
      \expandafter\def\csname LTb\endcsname{\color{black}}%
      \expandafter\def\csname LTa\endcsname{\color{black}}%
      \expandafter\def\csname LT0\endcsname{\color{black}}%
      \expandafter\def\csname LT1\endcsname{\color{black}}%
      \expandafter\def\csname LT2\endcsname{\color{black}}%
      \expandafter\def\csname LT3\endcsname{\color{black}}%
      \expandafter\def\csname LT4\endcsname{\color{black}}%
      \expandafter\def\csname LT5\endcsname{\color{black}}%
      \expandafter\def\csname LT6\endcsname{\color{black}}%
      \expandafter\def\csname LT7\endcsname{\color{black}}%
      \expandafter\def\csname LT8\endcsname{\color{black}}%
    \fi
  \fi
  \setlength{\unitlength}{0.0500bp}%
  \begin{picture}(3854.00,3288.00)%
    \gplgaddtomacro\gplbacktext{%
      \csname LTb\endcsname%
      \put(384,512){\makebox(0,0)[r]{\strut{}$700$}}%
      \csname LTb\endcsname%
      \put(384,1029){\makebox(0,0)[r]{\strut{}$750$}}%
      \csname LTb\endcsname%
      \put(384,1545){\makebox(0,0)[r]{\strut{}$800$}}%
      \csname LTb\endcsname%
      \put(384,2062){\makebox(0,0)[r]{\strut{}$850$}}%
      \csname LTb\endcsname%
      \put(384,2578){\makebox(0,0)[r]{\strut{}$900$}}%
      \csname LTb\endcsname%
      \put(384,3095){\makebox(0,0)[r]{\strut{}$950$}}%
      \csname LTb\endcsname%
      \put(962,352){\makebox(0,0){\strut{}$1990$}}%
      \csname LTb\endcsname%
      \put(1564,352){\makebox(0,0){\strut{}$1995$}}%
      \csname LTb\endcsname%
      \put(2166,352){\makebox(0,0){\strut{}$2000$}}%
      \csname LTb\endcsname%
      \put(2769,352){\makebox(0,0){\strut{}$2005$}}%
      \csname LTb\endcsname%
      \put(3371,352){\makebox(0,0){\strut{}$2010$}}%
      \put(16,1803){\rotatebox{-270}{\makebox(0,0){\strut{}lifetime (s)}}}%
      \put(2166,112){\makebox(0,0){\strut{}year of publication}}%
    }%
    \gplgaddtomacro\gplfronttext{%
      \csname LTb\endcsname%
      \put(3118,2952){\makebox(0,0)[r]{\strut{}largest storage time}}%
      \csname LTb\endcsname%
      \put(3118,2792){\makebox(0,0)[r]{\strut{}neutron lifetime}}%
      \csname LTb\endcsname%
      \put(3371,1480){\makebox(0,0){\strut{}\rotatebox{00}{\tiny Pichlmaier\cite{Pichlmaier10}}}}%
      \put(2769,2103){\makebox(0,0){\strut{}\rotatebox{00}{\tiny Serebrov\cite{Serebrov05}}}}%
      \put(2166,1136){\makebox(0,0){\strut{}\rotatebox{00}{\tiny Arzumanov\cite{Arzumanov00}}}}%
      \put(1323,1603){\makebox(0,0){\strut{}\rotatebox{00}{\tiny Mampe\cite{Mampe93}}}}%
      \put(1203,2180){\makebox(0,0){\strut{}\rotatebox{00}{\tiny Nesvizhevsky\cite{Nesvi92}}}}%
      \put(841,666){\makebox(0,0){\strut{}\rotatebox{00}{\tiny Mampe\cite{Mampe89}}}}%
    }%
    \gplbacktext
    \put(0,0){\includegraphics{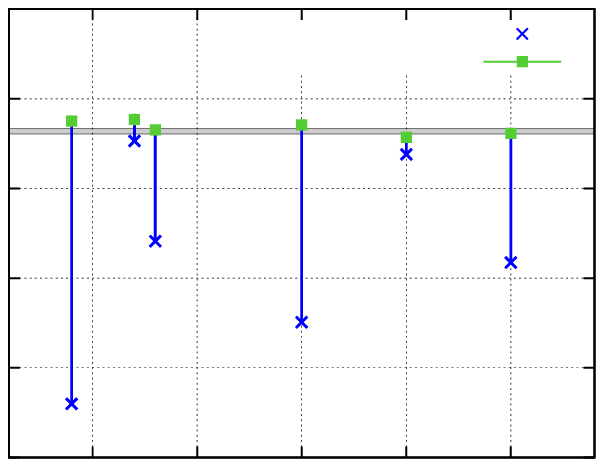}}%
    \gplfronttext
  \end{picture}%
\endgroup

%% file: lambda.tex
\begingroup
  \makeatletter
  \providecommand\color[2][]{%
    \GenericError{(gnuplot) \space\space\space\@spaces}{%
      Package color not loaded in conjunction with
      terminal option `colourtext'%
    }{See the gnuplot documentation for explanation.%
    }{Either use 'blacktext' in gnuplot or load the package
      color.sty in LaTeX.}%
    \renewcommand\color[2][]{}%
  }%
  \providecommand\includegraphics[2][]{%
    \GenericError{(gnuplot) \space\space\space\@spaces}{%
      Package graphicx or graphics not loaded%
    }{See the gnuplot documentation for explanation.%
    }{The gnuplot epslatex terminal needs graphicx.sty or graphics.sty.}%
    \renewcommand\includegraphics[2][]{}%
  }%
  \providecommand\rotatebox[2]{#2}%
  \@ifundefined{ifGPcolor}{%
    \newif\ifGPcolor
    \GPcolortrue
  }{}%
  \@ifundefined{ifGPblacktext}{%
    \newif\ifGPblacktext
    \GPblacktexttrue
  }{}%
  \let\gplgaddtomacro\g@addto@macro
  \gdef\gplbacktext{}%
  \gdef\gplfronttext{}%
  \makeatother
  \ifGPblacktext
    \def\colorrgb#1{}%
    \def\colorgray#1{}%
  \else
    \ifGPcolor
      \def\colorrgb#1{\color[rgb]{#1}}%
      \def\colorgray#1{\color[gray]{#1}}%
      \expandafter\def\csname LTw\endcsname{\color{white}}%
      \expandafter\def\csname LTb\endcsname{\color{black}}%
      \expandafter\def\csname LTa\endcsname{\color{black}}%
      \expandafter\def\csname LT0\endcsname{\color[rgb]{1,0,0}}%
      \expandafter\def\csname LT1\endcsname{\color[rgb]{0,1,0}}%
      \expandafter\def\csname LT2\endcsname{\color[rgb]{0,0,1}}%
      \expandafter\def\csname LT3\endcsname{\color[rgb]{1,0,1}}%
      \expandafter\def\csname LT4\endcsname{\color[rgb]{0,1,1}}%
      \expandafter\def\csname LT5\endcsname{\color[rgb]{1,1,0}}%
      \expandafter\def\csname LT6\endcsname{\color[rgb]{0,0,0}}%
      \expandafter\def\csname LT7\endcsname{\color[rgb]{1,0.3,0}}%
      \expandafter\def\csname LT8\endcsname{\color[rgb]{0.5,0.5,0.5}}%
    \else
      \def\colorrgb#1{\color{black}}%
      \def\colorgray#1{\color[gray]{#1}}%
      \expandafter\def\csname LTw\endcsname{\color{white}}%
      \expandafter\def\csname LTb\endcsname{\color{black}}%
      \expandafter\def\csname LTa\endcsname{\color{black}}%
      \expandafter\def\csname LT0\endcsname{\color{black}}%
      \expandafter\def\csname LT1\endcsname{\color{black}}%
      \expandafter\def\csname LT2\endcsname{\color{black}}%
      \expandafter\def\csname LT3\endcsname{\color{black}}%
      \expandafter\def\csname LT4\endcsname{\color{black}}%
      \expandafter\def\csname LT5\endcsname{\color{black}}%
      \expandafter\def\csname LT6\endcsname{\color{black}}%
      \expandafter\def\csname LT7\endcsname{\color{black}}%
      \expandafter\def\csname LT8\endcsname{\color{black}}%
    \fi
  \fi
  \setlength{\unitlength}{0.0500bp}%
  \begin{picture}(4988.00,3288.00)%
    \gplgaddtomacro\gplbacktext{%
      \csname LTb\endcsname%
      \put(384,512){\makebox(0,0)[r]{\strut{}$-1.290$}}%
      \csname LTb\endcsname%
      \put(384,835){\makebox(0,0)[r]{\strut{}$-1.285$}}%
      \csname LTb\endcsname%
      \put(384,1158){\makebox(0,0)[r]{\strut{}$-1.280$}}%
      \csname LTb\endcsname%
      \put(384,1481){\makebox(0,0)[r]{\strut{}$-1.275$}}%
      \csname LTb\endcsname%
      \put(384,1803){\makebox(0,0)[r]{\strut{}$-1.270$}}%
      \csname LTb\endcsname%
      \put(384,2126){\makebox(0,0)[r]{\strut{}$-1.265$}}%
      \csname LTb\endcsname%
      \put(384,2449){\makebox(0,0)[r]{\strut{}$-1.260$}}%
      \csname LTb\endcsname%
      \put(384,2772){\makebox(0,0)[r]{\strut{}$-1.255$}}%
      \csname LTb\endcsname%
      \put(384,3095){\makebox(0,0)[r]{\strut{}$-1.250$}}%
      \csname LTb\endcsname%
      \put(641,352){\makebox(0,0){\strut{}$1985$}}%
      \csname LTb\endcsname%
      \put(1446,352){\makebox(0,0){\strut{}$1990$}}%
      \csname LTb\endcsname%
      \put(2251,352){\makebox(0,0){\strut{}$1995$}}%
      \csname LTb\endcsname%
      \put(3055,352){\makebox(0,0){\strut{}$2000$}}%
      \csname LTb\endcsname%
      \put(3860,352){\makebox(0,0){\strut{}$2005$}}%
      \csname LTb\endcsname%
      \put(4665,352){\makebox(0,0){\strut{}$2010$}}%
      \put(-368,1803){\rotatebox{-270}{\makebox(0,0){\strut{}$\lambda = g_A/g_V$}}}%
      \put(2733,112){\makebox(0,0){\strut{}year of publication}}%
    }%
    \gplgaddtomacro\gplfronttext{%
      \csname LTb\endcsname%
      \put(2208,1135){\makebox(0,0)[r]{\strut{}our average}}%
      \csname LTb\endcsname%
      \put(2208,975){\makebox(0,0)[r]{\strut{}included in PDG 2010}}%
      \csname LTb\endcsname%
      \put(2208,815){\makebox(0,0)[r]{\strut{}\textsc{Perkeo II}}}%
      \csname LTb\endcsname%
      \put(2208,655){\makebox(0,0)[r]{\strut{}UCNA}}%
      \csname LTb\endcsname%
      \put(3377,1364){\makebox(0,0){\strut{}\rotatebox{00}{\tiny Abele\cite{Abele02}}}}%
      \put(3216,1529){\makebox(0,0){\strut{}\rotatebox{00}{\tiny Mostovoi\cite{Mostovoi01}}}}%
      \put(2653,1739){\makebox(0,0){\strut{}\rotatebox{00}{\tiny Liaud\cite{Liaud97}}}}%
      \put(2573,2178){\makebox(0,0){\strut{}\rotatebox{00}{\tiny Yerozilimsky\cite{Yero97}}}}%
      \put(802,1933){\makebox(0,0){\strut{}\rotatebox{00}{\tiny Bopp\cite{Bopp86}}}}%
      \put(4665,1067){\makebox(0,0){\strut{}\rotatebox{00}{\tiny Liu\cite{Liu10}}}}%
      \put(4343,1358){\makebox(0,0){\strut{}\rotatebox{00}{\tiny Abele\cite{Abele08}}}}%
    }%
    \gplbacktext
    \put(0,0){\includegraphics{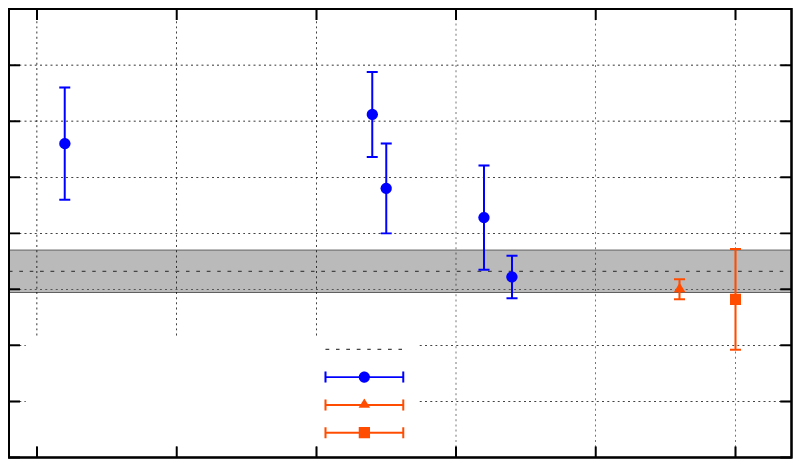}}%
    \gplfronttext
  \end{picture}%
\endgroup

%% file: ckm2010.bbl
\begin{thebibliography}{00}
\setlength{\itemsep}{-2mm}

\bibitem{Jackson57} J.\,D. Jackson, S.B. Treiman, and H.W. Wyld, Jr., Phys. Rev. 106, 517 (1957)
\bibitem{Wilkinson82} D.\,H. Wilkinson, Nucl. Phys. A377, 474 (1982)
\bibitem{Marciano06} W.\,J. Marciano and A. Sirlin,  Phys. Rev. Lett. 96, 032002 (2006)
\bibitem{Marciano10} W.\,J. Marciano, this conference

\bibitem{PDG2010} K. Nakamura et al. (Particle Data Group), J. Phys. G 37, 075021 (2010) 

\bibitem{Spivak88} P.\,E. Spivak, JETP 67, 1735 (1988)
\bibitem{Mampe89} W. Mampe \textit{et al.}, Phys. Rev. Lett. 63, 593 (1989)
\bibitem{Nesvi92} V.\,V. Nesvizhevsky \textit{et al.}, JETP 75, 405 (1992)
\bibitem{Mampe93} W. Mampe \textit{et al.}, JETP Lett. 57, 82 (1993)
\bibitem{Byrne96} J. Byrne \textit{et al.}, EPL 33, 187 (1996)
\bibitem{Arzumanov00} S. Arzumanov \textit{et al.}, Phys. Lett. B 483, 15 (2000)
\bibitem{Nico05} J.\,S. Nico \textit{et al.}, Phys. Rev. C 71, 055502 (2005)
\bibitem{Serebrov05} A. Serebrov \textit{et al.}, Phys. Lett. B 605, 72 (2005)
\bibitem{Pichlmaier10} A. Pichlmaier \textit{et al.}, Phys. Lett. B 693, Vol. 3, 221-226 (2010)
\bibitem{Serebrov10} A.\,P. Serebrov and A.\,K. Fomin, Phys. Rev. C 82, 035501 (2010)

\bibitem{Bopp86} P. Bopp \textit{et al.}, Phys. Rev. Lett. 56, 919 (1986)
\bibitem{Liaud97} P. Liaud \textit{et al.}, Nucl. Phys. A 612, 53 (1997)
\bibitem{Yero97} B.\,G. Yerozolimsky \textit{et al.}, Phys. Lett. B 412, 240 (1997)
\bibitem{Mostovoi01} Yu.\,A. Mostovoi \textit{et al.}, Phys.\ Atom.\ Nucl.\ 64, 1955 (2001)
\bibitem{Abele02} H. Abele \textit{et al.}, Phys. Rev. Lett. 88, 211801 (2002)
\bibitem{Abele08} H. Abele, Prog. Part. Nucl. Phys. 60, 1 (2008)
\bibitem{Liu10} J. Liu \textit{et al.}, Phys. Rev. Lett. 105, 181803 (2010) 
\bibitem{Mund11} D. Mund \textit{et al.}, in preparation

\bibitem{Paul09} S. Paul, Nucl. Instr. Meth. A 611 (2009)
\bibitem{Nico09} J.\,S. Nico, J. Phys. G 36 (2009)
\bibitem{Konrad10} G. Konrad \textit{et al.}, in: H.\,V. Klapdor-Kleingrothaus \emph{et al.} (eds.), World Scientific, ISBN 978-981-4340-85-4 (2011) and arXiv:1007.3027 (2010)
\bibitem{Dubbers11} D. Dubbers and M.\,G. Schmidt, Rev. Mod. Phys., in print (2011) and arXiv:1105.3694v1 [hep-ph]

\bibitem{Hardy09} J.\,D. Hardy, I. Towner, Phys. Rev. C 79, 055502 (2009)
\bibitem{Towner10} I. Towner, this conference
\bibitem{Malconian10} D. Melconian, this conference

\bibitem{Dubbers08} D. Dubbers \textit{et al.}, Nucl. Instr. Meth. A 596 (2008)
\bibitem{Maerkisch09} B. M{\"a}rkisch \textit{et al.}, Nucl. Instr. Meth. A 611 (2009) 
\bibitem{Soldner09} T. Solder \textit{et al.} (eds.), Nucl. Instr. Meth. A 611, Vol. 2-3, 111-344  (2009) 


\end{thebibliography}
